\documentclass[prl,twocolumn,amsmath,amssymb,superscriptaddress]{revtex4}

\usepackage{graphicx} 
\begin{document}

\newcommand{\ltwid}{\mathrel{\raise.3ex\hbox{$<$\kern-.75em\lower1ex\hbox{$\sim$}}}} 
\newcommand{\gtwid}{\mathrel{\raise.3ex\hbox{$>$\kern-.75em\lower1ex\hbox{$\sim$}}}} 
\def\K{{\bf{K}}} 
\def\Q{{\bf{Q}}} 
\def\Gbar{\bar{G}} 
\def\tk{\tilde{\bf{k}}} 
\def\k{{\bf{k}}} 
\def\kt{{\tilde{\bf{k}}}} 
\def\p{{\bf{p}}} 
\def\q{{\bf{q}}} 
\def\pp{{\bf{p}}^\prime} 
\def\Gpp{\Gamma^{pp}} 
\def\Phid{\Phi_d(\k,\omega_n)} 
\def\ld{\lambda_d(T)} 
\def\n{\langle n \rangle } 
\def\dw{d_{x^2-y^2}} 
\def\Ub{\bar{U}}

\title{Neutron scattering as a probe of the Fe-pnicitide superconducting gap}

\author{T.A.~Maier} \email{maierta@ornl.gov}  \affiliation{Center for Nanophase Materials Sciences and Computer Science and Mathematics Division, Oak Ridge National Laboratory, Oak Ridge, TN 37831-6164}

\author{D.J.~Scalapino} \email{djs@physics.ucsb.edu} \affiliation{Department of Physics, University of California, Santa Barbara, CA 93106-9530} 

\date{\today} 
\begin{abstract}
	Inelastic neutron scattering provides a probe for studying the spin and momentum structure of the superconducting gap. Here, using a two-orbital model for the Fe-pnicitide superconductors and an RPA-BCS approximation for the dynamic spin susceptibility, we explore the scattering response for various gaps that have been proposed.
\end{abstract}

\maketitle

\textit{Introduction - } 
Recent neutron scattering experiments have shown that LaOFeAs undergoes a structural distortion below $\sim$150K, which is then followed at $\sim$137K by the onset of long-range spin density wave (SDW) order with a wave vector $\q=(0.5,0.5,0.5)\pi/a$ \cite{dong-2008,dai:08}. When 
it is doped with F, both the structural distortion and the magnetic order are suppressed and the system becomes superconducting \cite{dai:08}. When La is replaced by Sm, superconducting transition temperatures of 55K have been reported \cite{ren-2008}. Thus it is natural to believe that the Fe-pnicitide superconductors have an electronic pairing mechanism and a variety of unconventional gap structures have already been proposed \cite{mazin:08,dai-2008,xu-2008,kuroki-2008,lee-2008,si-2008,yao-2008,qi-2008}. Here we explore how inelastic neutron scattering in the superconducting state can provide information on which of the gap structures actually occurs. 

Bandstructure calculations for doped LaOFeAs give a Fermi surface for the 2-Fe/cell Brillouin zone which consists of two hole cylinders around the $\Gamma$-point and two-electron cylinders around the M-point \cite{xu-2008,mazin:08,singh-2008}. When this is folded out into the larger Brillouin zone which is associated with a square lattice of Fe sites having 1-Fe/cell, one gets a similar Fermi surface to that shown in Fig.~1a \cite{mazin:08}. In this case the SDW would be associated with $\q=(\pi,0)$ or $(0,\pi)$. Here we will use a simple 2-orbital per site tight-binding model, which has been parametrized to give the Fermi surface shown in Fig.~1 \cite{raghu:08}. We then assume that the spin susceptibility in the superconducting state can be modeled by an RPA-BCS form and proceed to explore the structure of the inelastic scattering in the superconducting state for two $\q$ values and various gaps. As one knows from the cuprate problem, the occurance of resonances in the neutron scattering depend through the BCS coherence factors on the relative signs of the gap on different parts of the Fermi surface which are separated by $\q$. For the present case, in which there are multiple Fermi surfaces, there is a variety of ways in which resonances can occur and provide information on the gap structure. 

In the following, we first give a brief review of the model and then outline the RPA-BCS calculation of the spin susceptibility. This is similar to various approximations used for both the cuprates \cite{PhysRevLett.75.4134} and Sr$_2$RuO$_4$ \cite{PhysRevLett.86.5978}. We calculate the normal RPA spin susceptibility and then examine the RPA-BCS response in the superconducting state for both singlet and triplet gap functions. A related study based on a four-band model was reported in Ref.~\cite{korshunov-2008} for the case of a singlet gap. However, these authors neglected to take into account the matrix elements which relate the band operators to the orbital operators. The singlet gap functions we will use correspond to low order lattice harmonic representations of the sign-reversed s-wave gap proposed by Mazin {\it et al.} \cite{mazin:08},
\begin{equation} \label{eq:s*}
	\Delta_{s*}(\k) = \Delta_0 \cos k_x \cos k_y
\end{equation}
and an extended $s$-wave gap
\begin{equation} \label{eq:dxy}
	\Delta_{xs}(\k) = \Delta_0/2 (\cos k_x + \cos k_y)\,.
\end{equation}
We will aso determine the spin-flip ${\rm Im}\chi^{+-}(\q,\omega)$ and non-spin-flip ${\rm Im}\chi_{zz}(\q,\omega)$ response for various p-wave triplets. In this case,
\begin{equation} \label{eq:pw}
	\Delta_{\alpha\beta}(\k) = \left[{\vec d}(\k)\cdot {\vec \sigma}i\sigma_2\right]_{\alpha\beta}
\end{equation}
with $\vec{d}$ perpendicular to the Fe plane and $d_z(\k)=\Delta(\k)$. Here we will explore $\Delta(\k)=\sin k_x$, $\sin 2k_x$ and $\sin k_x + i \sin k_y$.

We conclude with a summary of what one can expect to learn about the gap symmetry from inelastic neutron scattering in the superconducting state. 


\textit{Model - } 
Bandstructure calculations for doped LaOFeAs show that the low energy states near the Fermi energy have dominant 3d Fe character and various multi-orbital tight-binding fits have been proposed. Here we will use a minimal, 2-orbital "$d_{xz}$-$d_{yz}$" per site tight-binding model with parameters chosen to give the Fermi surfaces shown in Fig.1a. This model has the virtue of simplicity while qualitatively capturing the shapes of the bandstructure Fermi surfaces and the relationship between the band operators to the orbital operators. This latter feature is important since it is the variation of the gaps on the Fermi surfaces that determine, through the BCS coherence factors, the inelastic neutron scattering response. While the magnitude of the response depends upon the Fermi velocities on the Fermi surfaces, which are not well reproduced by the 2-orbital model, the occurance or non-occurance of resonant features is determined by the k-dependence of the gap and the Femri surfaces. 

As described in Ref.~\cite{raghu:08}, our minimal model consists of a square two-dimensional lattice with degenerate "$d_{xz}$" and "$d_{yz}$" orbitals on each site. One-electron hopping parameters $t_i$ are introduced which provide near-neighbor $\sigma(t_1)$ and $\pi(t_2)$ couplings between similar orbitals, as well as a next near-neighbor coupling $t_3$. In addition, there is a next-near-neighbor coupling $t_4$ which hybridizes $d_{xz}$ with $d_{yz}$. The resulting tight-binding Hamiltonian can be written as 
\begin{equation} \label{eq:H0}
	H_0 = \sum_{\k\sigma} \psi^\dagger_{\sigma}(\k)\left[(\varepsilon_+(\k)-\mu)\mathbb{I}+\varepsilon_-(\k)\tau_3+\varepsilon_{xy}(\k)\tau_1\right]\psi^{\phantom\dagger}_\sigma
\end{equation}
with
\begin{equation} \label{eq:psi}
	\psi_\sigma(\k) = \left( \begin{array}{c} d_{x\sigma}(\k)\\d_{y\sigma}(\k)\end{array} \right)\,.
\end{equation}
Here $\tau_i$ are the usual Pauli matrices and 
\begin{eqnarray} \label{eq:eps}
	\varepsilon_{\pm}(\k) &=& \frac{1}{2}[\varepsilon_x(\k)\pm\varepsilon_y(\k)]\,,\nonumber\\
	\varepsilon_x(\k) &=& -2t_1 \cos k_x - 2t_2 \cos k_y - 4t_3 \cos k_x\cos k_y\,,\nonumber\\
	\varepsilon_y(\k) &=& -2t_2 \cos k_x - 2t_1 \cos k_y - 4t_3 \cos k_x\cos k_y\,,\nonumber\\
	\varepsilon_{xy}(\k) &=& -4t_4\sin k_x \sin k_y\,.
\end{eqnarray}
Introducing the band-operator $\gamma_{\nu\sigma}(\k)$, such that for $r=x$ or $y$
\begin{equation} \label{eq:psigamma}
	\psi_{r\sigma}(\k)= \sum_{\nu=\pm} a^\nu_r(\k) \gamma_{\nu\sigma}(\k)
\end{equation}
with 
\begin{eqnarray} \label{eq:knu}
	a_+^x(\k) &=& a_-^y(\k) = {\rm sgn}(\varepsilon_{xy}(\k))
 \sqrt{\frac{1}{2}+\frac{\varepsilon_-(\k)}{2\sqrt{\varepsilon^2_-(\k)+\varepsilon_{xy}^2(\k)}}}\nonumber\\
a_+^y(\k) &=& -a_-^x(\k) = 
 \sqrt{\frac{1}{2}-\frac{\varepsilon_-(\k)}{2\sqrt{\varepsilon^2_-(\k)+\varepsilon_{xy}^2(\k)}}}\,,
\end{eqnarray}
the tight-binding Hamiltonian becomes
\begin{equation} \label{eq:H0diag}
	H_0 = \sum_{\nu=\pm,\sigma} E_\nu(\k) \gamma_{\nu\sigma}^\dagger(\k)\gamma^{\phantom\dagger}_{\nu\sigma}(\k)\,.
\end{equation}
Here the band-energies are
\begin{equation} \label{eq:Epm}
	E_\pm(\k) = \varepsilon_+(\k) \pm \sqrt{\varepsilon_-^2(\k)+\varepsilon_{xy}^2(\k)}-\mu\,.
\end{equation}
We choose the hopping parameters $t_1=-1$, $t_2=1.3$, $t_3=t_4=-0.85$ and chemical potential $\mu=1.45$. 
The $\alpha_1$ and $\alpha_2$ Fermi surfaces in Fig.~1a correspond to $E_-(\k_F)=0$, while the $\beta_1$ and $\beta_2$ surfaces correspond to $E_+(\k_F)=0$.


\textit{Inelastic neutron scattering - } 
The physical spin susceptibility 
\begin{equation} \label{eq:chiphys}
	\chi^{ij}(q,i\omega_m) = \sum_{r,s}\chi^{ij}_{rs}(\q,i\omega_m)
\end{equation}
is calculated from the orbital dependent spin susceptibility defined as
\begin{equation} \label{eq:chiorb}
	\chi^{ij}_{rs}(\q,i\omega_m) = \int\limits_0^\beta d\tau \langle T_\tau S_i^r(\q,\tau) S_j^s(-\q,0) \rangle \,.
\end{equation}
Here, $r,s=x,y$ label the orbital indices, and $S_i^r(\q)=\frac{1}{2}\sum_\k\psi^\dagger_{r\alpha}(\k+\q)\sigma^i_{\alpha\beta}\psi_{r\beta}(\k)$ is the $i$th-component of the spin operator for orbital $r$. In the BCS framework, one obtains for the BCS orbital dependent spin susceptibility
\begin{eqnarray} \label{eq:chi0SC}	
	\chi^{ij}_{0,rs}(\q,\omega_m) = -\frac{1}{2}\sigma^i_{\alpha\beta}\sigma^j_{\gamma\delta} \sum_{\k,n} M^{\nu\nu'}_{rs}(\k,\q)\times\\
	&& \hspace{-6cm} \left\{ G_{\delta\alpha}^\nu(\k+\q)G_{\beta\gamma}^{\nu'}(\k) + F_{\alpha\gamma}^{\dagger\nu}(-\k-\q)F_{\beta\delta}^{\nu'}(\k) \right\}\nonumber\,.
\end{eqnarray}
Here we used $k=(\k,\omega_n)$ and $q=(\q,\omega_m)$ and $\nu,\nu'=+,-$ are the eigenvalues of the bands. The normal and anomalous Green's functions for band $\nu$ are given by 
\begin{equation} \label{eq:gandF}
	G_{\alpha\beta}^\nu(\k) = -\delta_{\alpha\beta}\frac{i\omega_m+E_\nu(\k)}{\omega_m^2+{\cal E}_\nu^2(\k)}\,,\,\, F^\nu_{\alpha\beta}(\k) = \frac{\Delta_{\alpha\beta}(\k)}{\omega_m^2+{\cal E}_\nu^2(\k)}\,,
\end{equation}
with ${\cal E}_\nu(\k) = \sqrt{E_\nu^2(\k)+|\Delta(\k)|^2}$. The hybridization between the bands is reflected in the matrix elements
\begin{equation} \label{eq:Mmatrix}
	M_{rs}^{\nu\nu'}(\k,\q)=a^r_\nu(\k+\q)a^r_{\nu'}(\k)a^s_{\nu'}(\k)a^s_{\nu}(\k+\q)\,.
\end{equation}

One can then obtain the BCS spin susceptibilty $\chi^{ij}_0(\q,\omega)=\sum_{r,s} \chi^{ij}_{0,rs}(\q,\omega)$ on the real frequency axis from an analytical continuation of Eq.~(\ref{eq:chi0SC}).
We then use the RPA to take into account the effect of the on-site intra-orbital Coulomb interactions $U$ \cite{raghu:08}. The RPA susceptibility is determined from the matrix equation
\begin{equation} \label{eq:chiRPA}
	\chi^{ij}_{\rm RPA}(q) = \sum_{r,s} \left[\chi^{ij}_0(q)\left( \mathbb{I}-\Gamma\chi^{ij}_0(q) \right)^{-1}\right]_{rs}\,,
\end{equation}
where $q=(\q,\omega)$,
with the interaction vertex
\begin{equation} \label{eq:Gamma}
	\Gamma = \left( \begin{array}{cc} U & 0\\ 0 & U\end{array} \right)\,.
\end{equation}

\begin{figure} 
	[htbp] \centering
	\includegraphics[width=3.5in]{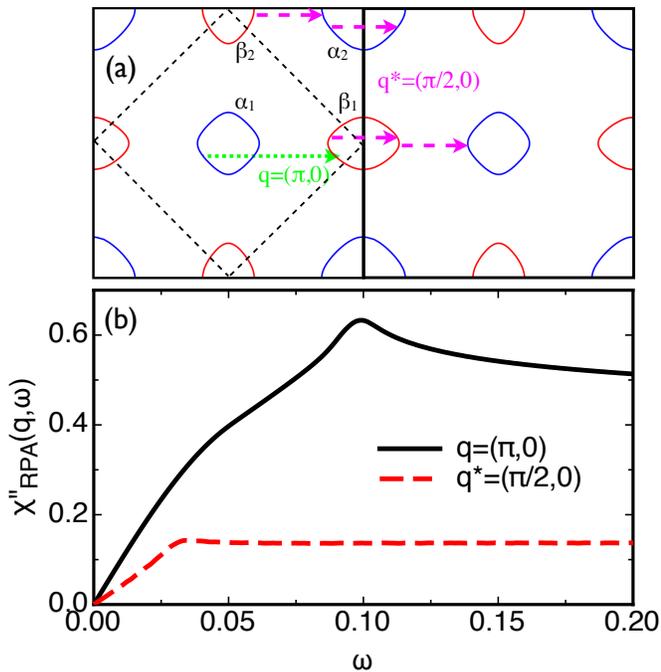}
	\caption{(a) Fermi surface of the two-orbital model in the large 1-Fe/cell Brillouin zone. When this is folded down along the dashed line, one obtains the Fermi surface of the 2-Fe/cell system. The $\alpha_{1/2}$ Fermi surface sheets are hole pockets given by $E_-(\k_F)=0$ and the $\beta_{1/2}$ sheets are electron pockets given by $E_+(\k_F)=0$. The nesting vectors $\q=(\pi,0)$ and $\q*=(\pi/2,0)$ are indicated by the dotted and dashed lines, respectively. (b) RPA spin susceptibility $\chi''_{\rm RPA}(\q,\omega)$ versus frequency in the normal state ($\Delta(\k)=0$) for $\q=(\pi,0)$ and $\q*=(\pi/2,0)$.}
	\label{fig:1}
\end{figure}

Fig.~\ref{fig:1}b shows the results for the imaginary part of the RPA spin susceptiblity $\chi^{+-}_{0,{\rm RPA}}(\q,\omega)$ in the normal state ($\Delta_0=0$) for $\q=(\pi,0)$ and $\q*=(\pi/2,0)$. Here we used $U=3|t_1|$, $\mu=1.45$ and the temperature $T=0.001|t_1|$.

For $\mu=1.45$, the static RPA spin susceptibility shows peaks at $\q=(\pi,0)$ and $\q*=(\pi/2,0)$. As indicated by the dotted and dashed arrows in Fig.~1a, $\q=(\pi,0)$ is a nesting vector between a region on the $\alpha_1$ Fermi surface sheet with the $\beta_1$ Fermi surface sheet, while $\q*=(\pi/2,0)$ connects several different regions on the Fermi surface. The scattering for $\q*=(\pi/2,0)$ is, however, dominated by the intra-band excitations on the $\beta_1$ Fermi surface sheet (see Fig.~1b), since for the other inter-band processes $\varepsilon_{xy}(\k)$ and $\varepsilon_{xy}(\k+\q*)$ are close to zero, and therefore the matrix-elements $M_{rs}^{\nu\nu'}(\k,\q)$ are very small. At low frequency $\chi''_{\rm RPA}(\q,\omega)$ is larger for $\q=(\pi,0)$ than for $\q=(\pi/2,0)$, since larger regions of the Fermi surface are nested for $\q=(\pi,0)$. At low frequency the RPA susceptibility is strongly enhanced over the unrenormalized susceptibility. 

In the superconducting state, the gap $\Delta(\k)$ is finite and the susceptibility, Eq.~(\ref{eq:chi0SC}), depends on the symmetry of the gap. For singlet pairing one has $F_{\uparrow\downarrow}(\k) = -F_{\downarrow\uparrow}(\k)$ and $F_{\uparrow\downarrow}(-\k) =  F_{\uparrow\downarrow}(\k)$ and hence the in-plane susceptibility $\chi^{+-} = \frac{1}{2}(\chi^{xx}+\chi^{yy})$ is equal to the out-of-plane susceptibility $\chi^{zz}$ as in the normal state. For the triplet case, however, one has 
$F_{\uparrow\downarrow}(\k) = F_{\downarrow\uparrow}(\k)$ and $F_{\uparrow\downarrow}(-\k) =  -F_{\uparrow\downarrow}(\k)$.Therefore $\chi^{+-}$ and $\chi^{zz}$ differ with respect to their superconducting coherence factors.


\textit{Results - } 
As is well known, the BCS coherence factors that enter the spin susceptibility depend upon the sign of $\Delta(\k+\q)\Delta(\k)$. For a singlet gap, when this is negative, there can be a resonance response at $\omega=|\Delta(\k+\q)|+|\Delta(\k)|$. For a triplet gap, the coherence factors are different for $\chi_{zz}$ and $\chi_{+-}$. In this case, when ${\rm Re}(\Delta^*(\k+\q)\Delta(\k))$ is negative, there can be a resonance in $\chi_{zz}$ but not in $\chi_{+-}$. Likewise, when  ${\rm Re}(\Delta^*(\k+\q)\Delta(\k))$ is positive, $\chi_{+-}$ can exhibit a resonance while $\chi_{zz}$ varies smoothly through $\omega=|\Delta(\k+\q)|+|\Delta(\k)|$. Thus, the neutron scattering response in the superconducting state can provide information on the momentum and spin structure of the superconducting gap. Here we examine the response for various gaps that have been proposed for LaOFeAs.

\begin{figure} [htbp] \centering
	\includegraphics[width=3.5in]{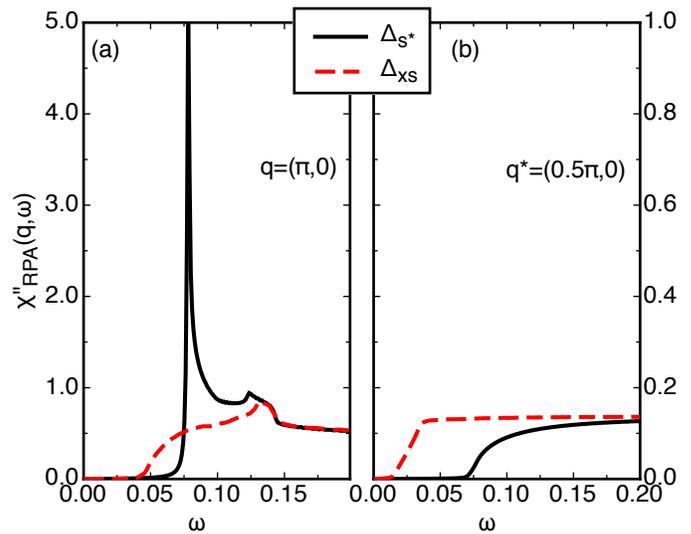}
	\caption{RPA dynamic spin susceptibility $\chi''_{\rm RPA}(\q,\omega)$ versus frequency in the spin singlet superconducting state for (a) $\q=(\pi,0)$, and (b) $\q*=(\pi/2,0)$.}
	\label{fig:2}
\end{figure}

To begin, Fig.~2a shows the imaginary part of the RPA-BCS spin susceptibility for various singlet gaps at a momentum transfer $\q=(\pi,0)$. We have modeled a sign-reversed s-wave gap $\Delta_{s*}(\k)$, i.e. the type proposed by Mazin {\it et al.} by Eq.~(\ref{eq:s*}) and the solid line in Fig.~\ref{fig:2}a shows the expected resonance response associated with having $\Delta(\k+\q)\Delta(\k)<0$. This behavior can be contrasted with the response found for an extended s-wave gap. Here $\Delta(\k+\q)\Delta(\k)>0$ for the $\q=(\pi,0)$ nesting vector. 

Fig.~2b shows the results for $\q*=(\pi/2,0)$. For this momentum transfer the scattering is dominated by the intra-band process connecting regions on the $\beta_1$ Fermi surface sheet. For these regions one has $\k+\q*=-\k$, and therefore $\Delta(\k+\q*)\Delta(\k)>0$ for the singlet gaps. Thus, no resonance is found for $\q*=(\pi/2,0)$ in this case.

\begin{figure} [htbp] \centering
	\includegraphics[width=3.5in]{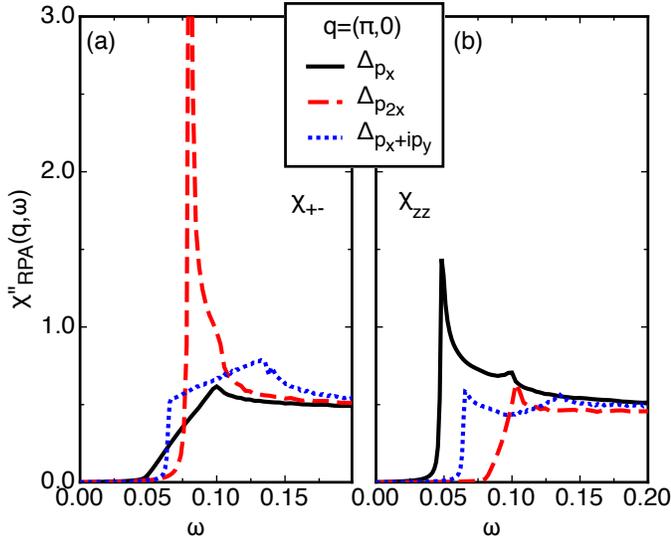}
	\caption{RPA  spin susceptibility (a) $\chi''_{+-,\rm RPA}(\q,\omega)$ and (b) $\chi''_{zz,\rm RPA}(\q,\omega)$ versus frequency in the spin triplet superconducting  state for $\q=(\pi,0)$.}
	\label{fig:3}
\end{figure}

Similar results for various triplet gaps are shown in Fig.~\ref{fig:3} for $\q=(\pi,0)$. For the $p_x$-wave $\sin k_x$ gap, $\Delta(\k+\q)\Delta(\k)<0$ and a resonance is seen in $\chi''_{zz}(\q,\omega)$ but not in $\chi''_{+-}(\q,\omega)$. One could also consider a sign-reversed p-wave modeled by $\Delta_{p2x}=\Delta_0\sin 2k_x$. Here the coherence factor for $\q=(\pi,0)$ is positive so that the resonance appears in $\chi''_{+-}(\q,\omega)$. For the $\sin k_x+i\sin k_y$ gap, $\sin k_x\sin(k_x+q_x)<0$ and $\sin k_y\sin(k_y+q_y)>0$ with similar size for the dominant process. Hence, ${\rm Re}(\Delta^*(\k+\q)\Delta(\k))$ is close to zero and one obtains qualitatively similar results for $\chi_{+-}$ and $\chi_{zz}$.  
 
\begin{figure} [htbp] \centering
	\includegraphics[width=3.5in]{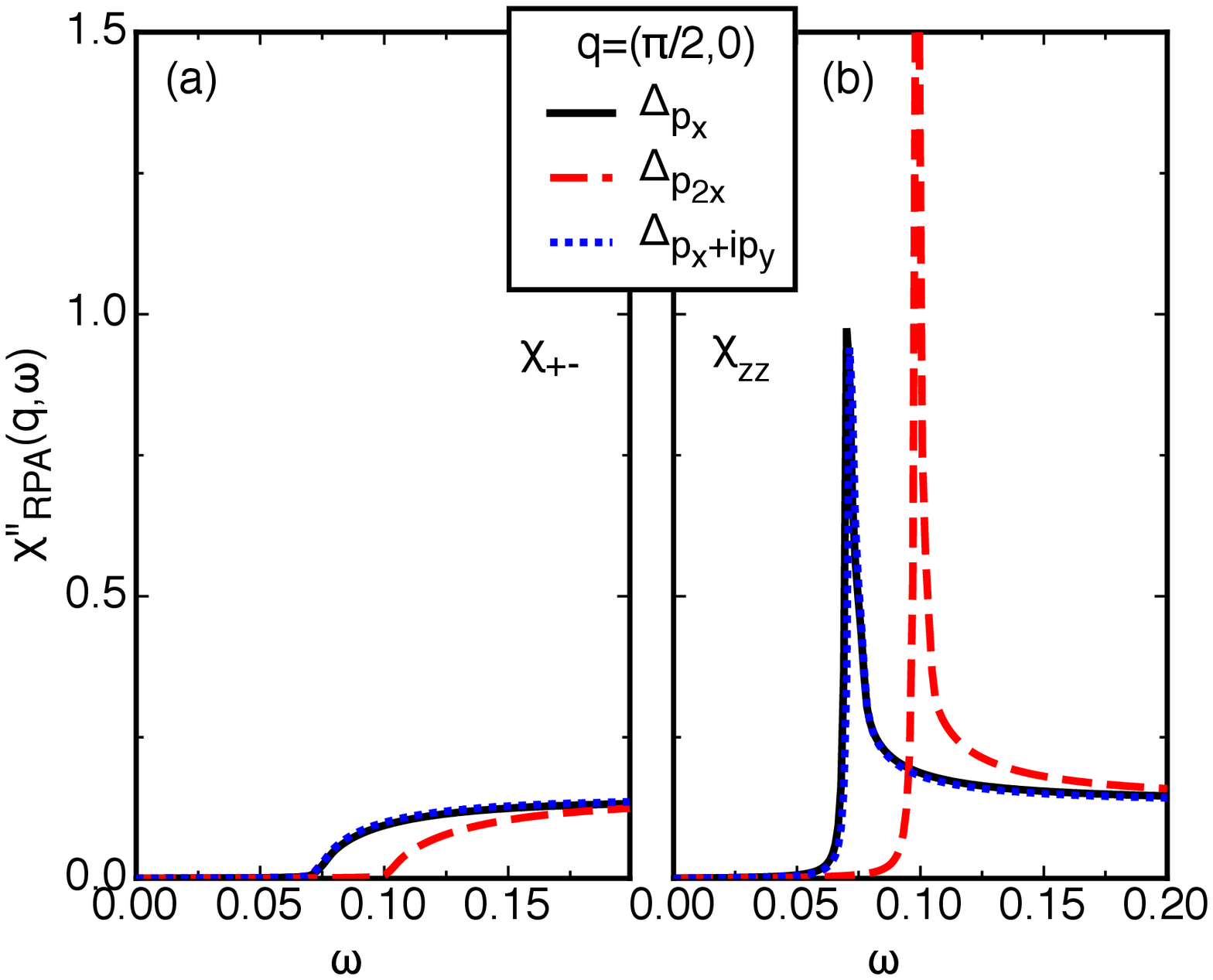}
	\caption{RPA  spin susceptibility (a) $\chi''_{+-,\rm RPA}(\q,\omega)$ and (b) $\chi''_{zz,\rm RPA}(\q,\omega)$ versus frequency in the spin triplet superconducting  state for  $\q*=(\pi/2,0)$.}
	\label{fig:4}
\end{figure}

Fig.~4 shows the results for the triplet gaps for $\q*=(\pi/2,0)$. For the $p_x$-wave $\sin k_x$ gap, the results are similar as in Fig.~3 for $\q=(\pi,0)$.  The gap changes sign under the transformation $\k\rightarrow -\k$ so that the gap has opposite signs on the sheets of the $\beta_1$ Fermi pocket connected by $\q*$ (see Fig.1a). Hence, $\chi''_{zz}(\q,\omega)$ displays a resonance while $\chi''_{+-}(\q,\omega)$ does not. For the $\sin 2k_x$ gap, the situation is similar and a resonance is found in $\chi_{zz}$, but not in $\chi_{+-}$. This is opposite to the results in Fig.3 for $\q=(\pi,0)$. The results for the $\sin k_x+i\sin k_y$ case are almost identical to the results found for the $p_x$-wave gap. This is explained by the fact that for the dominant intra-band scattering on the $\beta_1$ Fermi surface sheet, $\k_y$ and $\k_y+\q_y$ is close to zero, and therefore the $\sin k_y$ contribution to the  $p_x$+i$p_y$ gap is an order of magnitude smaller than the $\sin k_x$ contribution. 


\textit{Conclusion - } 
Using a two-orbital model for the Fe-pnicitide superconductors and an RPA-BCS approximation for the dynamic spin susceptibiltiy we have explored the inelastic scattering response for various gaps that have been proposed. As one would expect, we have found that the occurance of resonances in the dynamic spin susceptibility depends on the relative signs of the gap on the parts of the Fermi surface separated by $\q$, and in the case of triplet gaps also on whether the in-plane, $\chi''_{+-}(\q,\omega)$ or out-of-plane, $\chi''_{zz}(\q,\omega)$ components are studied. Specifically, for the singlet gaps, we have found that $\chi''(\q,\omega)$ displays a resonance for a sign-reversed s-wave gap modeled by $\Delta_0\cos k_x\cos k_y$ for $\q=(\pi,0)$ which corresponds to the antiferromagnetic wave-vector, but no resonance for the extended s-wave gap or for $\q*=(\pi/2,0)$. For the triplet case, resonances were found for a $p_x$-wave gap in $\chi''_{zz}$ and for a sign-reversed $p$-wave gap in $\chi''_{+-}$ for $\q=(\pi,0)$. For $\q*=(\pi/2,0)$, resonances appeared in $\chi''_{zz}$ for the $p$-wave, sign-reversed $p$-wave and $p_x$+i$p_y$-wave gaps. 

\textit{Acknowledgments - }
We acknowledge helpful discussions with I. Mazin, D. Singh, T.C. Schulthess S. Raghu and X.-L. Qi. A portion of this research at Oak Ridge National Laboratory's Center for Nanophase Materials Sciences was sponsored by the Scientific User Facilities Division, Office of Basic Energy Sciences, U.S. Department of Energy.


\bibliography{FeSC}


\end{document}